\documentclass[11pt]{article}

\usepackage[utf8]{inputenc}        
\usepackage[T1]{fontenc}           
\usepackage{parskip}               
\usepackage{setspace}              
\usepackage{fancyhdr}              
\usepackage{microtype}             
\usepackage[                       
  top=2.5cm,
  bottom=2.5cm,
  left=3cm,
  right=3cm
]{geometry} 

\usepackage{amsmath, amssymb, amsfonts}  
\usepackage{mathtools}                   
\usepackage{physics}                     
\usepackage{siunitx}                     
\usepackage{cancel}                      
\usepackage{bm}                          
\usepackage{esint}                       

\usepackage{graphicx}                   
\usepackage{wrapfig}                    
\usepackage{subcaption}                 
\usepackage{float}                      

\usepackage{tabularx}                  
\usepackage{booktabs}                  
\usepackage{longtable}                 
\usepackage{multirow}                  
\usepackage{array}                     

\usepackage{listings}                 
\usepackage{algorithm}                
\usepackage{algpseudocode}            

\usepackage{enumitem}                 
\usepackage{multicol}                 

\usepackage{xcolor}                   
\usepackage{todonotes}                

\usepackage{hyperref}                 
\hypersetup{
    colorlinks=true,
    linkcolor=blue,
    urlcolor=blue,
    citecolor=blue
}
\usepackage{cleveref}                 
\usepackage{pdfpages}                 

\usepackage[backend=biber,style=phys]{biblatex} 
\usepackage{tikz}
\usetikzlibrary{arrows.meta,positioning,decorations.pathmorphing,patterns}

\title{Exact Thermoelectric Transport Coefficients and Figure of Merit for
Graphene Photothermoelectric Devices from a Finite Zeta-Function Mott Series}

\author{
  Luis Daniel Villa Cortés\,\textsuperscript{1}
  \quad
  S.~R.~Valluri\,\textsuperscript{2}
  \quad
  Atul Jhalani\,\textsuperscript{3}
  \\[2pt]
  Ajay Soni\,\textsuperscript{3}
  \quad
  P.~C.~Deshmukh\,\textsuperscript{4}
  \\[8pt]
  {\small\itshape\textsuperscript{1}Universidad de Sonora, Hermosillo, Sonora, Mexico}\\
  {\small\itshape\textsuperscript{2}University of Western Ontario, London, Ontario, Canada}\\
  {\small\itshape\textsuperscript{3}Indian Institute of Technology Mandi, Mandi, Himachal Pradesh, India}\\
  {\small\itshape\textsuperscript{4}Central University of Kerala, Kasaragod, Kerala, India}\\[6pt]
  {\small\texttt{luis.daniel.villa.cortes@gmail.com}\quad
   \texttt{valluri@uwo.ca}\quad
   \texttt{atuljhalani27@gmail.com}}\\
  {\small\texttt{ajay@iitmandi.ac.in}\quad 
   \texttt{pcd@cukerala.ac.in}}
}

\date{\today}

\begin{document}
    \maketitle

    \begin{abstract}
	The standard Mott formula is widely used to describe thermoelectric transport, but it becomes less accurate when the temperature is not much smaller than the Fermi energy. In this work, we develop an all-orders extension of the Mott approach using a series of Riemann zeta functions. We show that when the transport function is a polynomial, the series ends after a finite number of terms, giving exact results within the model. We apply this method to graphene photothermoelectric devices using a quadratic conductivity model. The results provide closed-form expressions for the Seebeck coefficient, Lorenz ratio, and electronic figure of merit. The analysis shows that the Seebeck coefficient reaches a maximum instead of increasing indefinitely, while the Wiedemann–Franz law can be significantly violated at higher temperatures. We also find that disorder reduces the thermoelectric performance and that the electronic figure of merit has an upper limit in the clean graphene model. Finally, we discuss the effect of radiative heat transport on the figure of merit. These results provide a simple analytical way to study graphene thermoelectric transport beyond the usual low-temperature Mott approximation.
    \end{abstract}


\section{The zeta-function Mott series}

    The Mott formula relates the thermopower of a conductor to the energy derivative of its electrical conductivity at the Fermi level. It was popularized in the study of disordered systems by Cutler and Mott \cite{PhysRev.181.1336}, and Jonson and Mahan later showed that it becomes exact, to leading order in temperature, for independent electrons interacting with static impurities and with phonons treated in the adiabatic approximation \cite{jonson_mahan_1980}. Because of its simplicity, the formula is used routinely to interpret thermoelectric measurements, from bulk thermoelectric materials \cite{snyder_complex_2008,goldsmid_introduction_2016} to two-dimensional systems such as graphene, where early thermopower experiments \cite{zuev_thermoelectric_2009,wei_anomalous_2009} were analyzed precisely in terms of the Mott expression, and where its regime of validity has been examined theoretically \cite{hwang_thermopower_2009}. 
    
    In bulk thermoelectrics, raising the power factor entering the Mott formula is itself an active materials-design problem; one competing strategy is band-convergence engineering, e.g.\ via rare-earth doping in multiband, non-Dirac semiconductors such as SnTe \cite{acharya_rare_2018}, a materials-side approach unrelated to the graphene photothermoelectric framework developed here. However, as we show in this work, the standard formula is only the first term of an expansion in powers of $k_BT$ (where $k_B$ is the Boltzmann constant and $T$ is the absolute temperature), and it is known to fail when the temperature is not small compared with the chemical potential. In the specific context of graphene photothermoelectric devices, Antidormi and Cummings observed that the Mott formula severely overestimates the maximal Seebeck coefficient of high-mobility samples \cite{antidormi_optimizing_2021}. Related efforts to go beyond the leading-order expression, based on exact evaluations of the Fermi--Dirac integrals in terms of polylogarithms and the Lambert $W$ function, have been carried out in Refs.~\cite{yadav_analytic_2019,nair_modeling_2026}, building on the mathematical framework of Refs.~\cite{corless_lambertw_1996,valluri_applications_2000,valluri_quantum_2009}. The same functionhas recently been applied to graphene nanoribbons in a different context, where the branch-point structure of $W$ controls the sensitivity of quantum--confinement--based sensors \cite{chishtie_rineng_2026}.Riemann zeta constants have also surfaced in a Sommerfeld-like expansion elsewhere in thermoelectric transport: for resonant scattering in a disordered one-dimensional Landauer--B\"uttiker conductor, M\"uller, Smit \& Sigrist expand the conductance moments in the width of a resonance and find the leading coefficient fixed by the ratio $-\zeta(3/2)/(2\sqrt{2}\,\zeta(1/2))\approx0.6325$ \cite{muller_resonant_2015}, a structurally different but suggestive precedent for zeta-function constants appearing in this class of problem. In this section we derive the all-orders generalization of the Mott formula as a series of Riemann zeta functions.

    In the current treatments for this problem, three strategies are currently used to obtain thermoelectric coefficients beyond the leading Mott order, and each leaves a specific gap. (i) \emph{Truncated Sommerfeld expansions.} One keeps the $\mathcal{O}(k_BT/\mu)^2$ correction and stops. This is simple and analytic, but the expansion is asymptotic, not convergent, so adding terms eventually degrades the answer and no truncation order carries a rigorous error bound at the temperatures of interest. (ii) \emph{Exact Fermi--Dirac integrals in terms of polylogarithms and Lambert $W$ functions} \cite{yadav_analytic_2019,nair_modeling_2026}. This route is exact and has produced generalizations of the Wiedemann--Franz law and of $ZT$; its cost is that the resulting expressions involve special functions of a reduced chemical potential that must itself be obtained by inverting a transcendental equation, so optima are located numerically or branch by branch rather than in elementary closed form. (iii) \emph{Direct numerical quadrature} of the transport integrals, as done for graphene photothermoelectric devices by Antidormi and Cummings \cite{antidormi_optimizing_2021} and, for general band structures, by Boltzmann-transport codes such as BoltzTraP2 \cite{madsen_boltztrap2_2018}. This is accurate but yields no formula: parametric trends, bounds and optima must be re-computed sample by sample and cannot be read off or differentiated analytically.

    In this work, we resume the Sommerfeld expansion to all orders into a single indexed series of Riemann zeta functions, and we identify a condition (polynomial transport function) under which that series terminates. When it does, the result is not an improved approximation but an algebraic identity. The transport coefficients are exact and non-perturbative in $k_BT/\mu$ (where $\mu$ is the chemical potential and $k_B$ is the Boltzman constant), expressed in elementary functions, at every temperature for which the underlying conductivity model applies. This combines the analytic transparency of route (i) with the exactness of routes (ii) and (iii), and it does so without special functions or numerical inversion. The practical payoff appears in Section 2, where the standard graphene photothermoelectric conductivity is a finite polynomial and therefore falls exactly into this classand therefore the Seebeck coefficient, the Lorenz ratio and the electronic figure of merit collapse onto closed forms in only two dimensionless parameters. Because these forms are elementary, they can be differentiated, so the optima and bounds that route (iii) can only sample numerically are here obtained analytically, including a saturation mechanism for $|S|$ that explains, in closed form, the overestimation reported numerically in Ref.~\cite{antidormi_optimizing_2021}, a disorder threshold $l_c=1/4$ separating two qualitatively different Wiedemann--Franz violations, and a parameter-free ceiling on the electronic figure of merit of the model.

    \subsection{The transport function and the linearized Boltzmann picture}

    The typical Mott-Formula that appears in articles is \cite{zhao2025photonic}

    \begin{equation}
        S = -\frac{\pi^2k_B^2T}{3e}\,\frac{d\,\ln{\sigma}}{dE}\big{|}_{E=E_F}.
    \end{equation}

    Here $S$ is the Seebeck coefficient, $e$ is the fundamental charge, $E$ is the energy, $E_F$ is the Fermi energy, $k_B$ is the Boltzman constant, $T$ is the absolute temperature and $\sigma$ is the electrical conductivity. We'll follow the steps to derive a more generalized Mott-Formula. A form of $\sigma$ for independent carriers in the relaxation-time picture is \cite{AshcroftMermin1976,PhysRev.181.1336}

    \begin{equation} \label{generic_conductivity}
        \sigma = -e^2 \int_{-\infty}^{\infty}\frac{1}{3}\tau v^2\left(\frac{df}{dE}\right)N(E)dE.
    \end{equation}

    Here, $N(E)$ is the density of states, $\tau$ is the scattering time, $v$ is the carrier velocity and $f$ is the Fermi-Dirac function, which has the form \cite{Neamen2012SemiconductorPhysics} $f = 1/(\exp[(E-\mu)/k_BT]+1)$, where $\mu$ is the chemical potential. The factor $1/3$ in Eq.~(\ref{generic_conductivity}) arises from the isotropic angular average of $v_x^2$ over the Fermi surface in three dimensions; for a general dimensionality $d$ it is replaced by $1/d$, which is the form we adopt below. Physically, Eq.~(\ref{generic_conductivity}) states that only carriers within a thermal window around the chemical potential contribute to the current, each weighted by its diffusivity $\tau v^2/d$ and by the number of available states $N(E)$. By doing simpe math, we can see that

    \begin{equation}
        \frac{df}{dE} = -\frac{1}{k_BT} \, \frac{\exp[(E-\mu)/k_BT]}{(\exp[(E-\mu)/k_BT]+1)^2} = -\frac{1}{k_BT}f(1-f).
    \end{equation}

    Also, the scattering distance can be written as \cite{AshcroftMermin1976} $\Lambda^2 = \frac{1}{d}(\tau v)^2$, where $d$ is the number of space dimensions (for example, $d=2$ for graphene). So, let's rewrite (\ref{generic_conductivity}) as following:
    
    \begin{equation} \label{conductivity_2}
        \sigma = \int_{-\infty}^{\infty} \frac{e^2\Lambda^2}{\tau k_BT}\,f(1-f)  N(E) \, dE.
    \end{equation}

    In order to simplify the procedure, let's define the \textit{transport function} $G(E) \equiv \frac{e^2\Lambda^2}{\tau}N(E)$. So, (\ref{conductivity_2}) is now in terms of $G(E)$

    \begin{equation} \label{conductivity_3}
        \sigma = \frac{1}{k_BT}\int_{-\infty}^{\infty} f(1-f) G(E) \, dE.
    \end{equation}

    The transport function $G(E)$ (also called the transport distribution function) collects all the microscopic, material-specific information---band structure, scattering mechanisms and dimensionality---into a single function of energy. This is the same object whose shape Mahan and Sofo optimized in their classic analysis of the best possible thermoelectric \cite{mahan_sofo_1996}, and it is the quantity computed numerically by modern Boltzmann-transport packages such as BoltzTraP2 \cite{madsen_boltztrap2_2018}. Writing all transport coefficients in terms of $G(E)$ therefore keeps the derivation completely general, because nothing in what follows depends on the microscopic origin of $G$.

    \subsection{Moment expansion and the $M_{2i}$ integrals}

    We can see that $f(1-f)$ sharply peaked near the Chemical Potential $\mu$. This means that only electrons within an energy interval of order $k_BT$ around $\mu$ can efficiently contribute to transport. In other words, the kernel $f(1-f)$ is even in $E-\mu$ and peak with width of $\sim k_BT$. So, near $\mu$, the slowly varying part of the integrand may be expanded in powers of $E-\mu$. By making a Taylor expansion of $G(E)$, then (\ref{conductivity_3}) can be written as

    \begin{equation}
        \sigma = \frac{1}{k_BT}  \int_{-\infty}^{\infty} \sum_{i=0}^{\infty}{ \frac{G^{(i)}(\mu)}{i!}(E-\mu)^{i}f(1-f) \, dE}.
    \end{equation}

    Now, we assume that $G(E)$ is analytic in a neighborhood of $\mu$ and that its Taylor series converges absolutely against the weight $f(1-f)$. Then, by the Fubini--Tonelli theorem \cite{folland_real_1999}, the order of summation and integration may be interchanged.

    \begin{equation} \label{conductivity_4}
        \sigma = \frac{1}{k_BT} \sum_{i=0}^{\infty}{ \frac{G^{(i)}(\mu)}{i!} \int_{-\infty}^{\infty} (E-\mu)^{i}f(1-f) \, dE}.
    \end{equation}
    
    This step is the standard Sommerfeld-type expansion of degenerate-electron physics \cite{AshcroftMermin1976}. Two remarks are in order. First, for a generic (non-polynomial) transport function the resulting series is asymptotic rather than convergent, and its usefulness rests on the smallness of $k_BT/\mu$. Second if $G(E)$ is a polynomial in $E$, all derivatives above the polynomial degree vanish identically, the sum contains a finite number of terms, and the expansion is exact at any temperature.
    
    Because $f(1-f)$ is even, the odd terms vanish and we are only left with the even terms.

    \begin{equation} \label{conductivity_5}
        \sigma = \frac{1}{k_BT} \sum_{i=0}^{\infty}{ \frac{G^{(2i)}(\mu)}{(2i)!} \int_{-\infty}^{\infty} (E-\mu)^{2i}f(1-f) \, dE}.
    \end{equation}

    Now let's focus on the integral. Let's define

    \begin{equation}
        I_{2i} \equiv \int_{-\infty}^{\infty} (E-\mu)^{2i}f(1-f) \, dE.
    \end{equation}

    By making the transformation $x=(E-\mu)/k_BT$ and then $dE=k_BT\,dx$. We are left with

    \begin{equation}
        I_{2i} = (k_BT)^{2i+1} \int_{-\infty}^{\infty} x^{2i}\frac{e^x}{(e^x+1)^2} \, dx.
    \end{equation}

    Then, we can write \ref{conductivity_5} as
    \begin{equation}
        \sigma = \sum_{i=0}^{\infty}{ (k_BT)^{2i}\,\frac{G^{(2i)}(\mu)}{(2i)!} \int_{-\infty}^{\infty} x^{2i}\frac{e^x}{(e^x+1)^2} \, dx}.
    \end{equation}

    The change of variables leaves us with a mathematical problem. All the temperature dependence has been extracted as the prefactor $(k_BT)^{2i+1}$, and what remains is a family of universal dimensionless numbers (the moments of the thermal broadening kernel $e^x/(e^x+1)^2$). The integral is a dimensionless and even function. We'll focus on solving that problem. Let's define then

    \begin{equation}
        M_{2i} \equiv \int_{-\infty}^{\infty} x^{2i}\frac{e^x}{(e^x+1)^2} \, dx.
    \end{equation}

    For $i=0$ the result is trivial: $M_0=1$. So, we'll solve the problem for $i\geq1$. Because the integral is even, we can rewrite as following.

    \begin{equation} \label{M_function}
        M_{2i} = 2 \int_{0}^{\infty} x^{2i}\frac{e^x}{(e^x+1)^2} \, dx \:\: , \:\: i\geq1.
    \end{equation}

    It is easy to prove that $e^x/(e^x+1)^2 = e^{-x}/(e^{-x}+1)^2$. Then, results convenient to use the expansion $q/(q+1)^2 = \sum_{n=1}^{\infty}{(-1)^{n-1}\,n\,q^{n}}$ and choose $q=e^{-x}$. 

    \begin{equation}
        M_{2i} = 2 \int_{0}^{\infty} x^{2i}\frac{e^{-x}}{(e^{-x}+1)^2} \, dx = 2 \int_{0}^{\infty} x^{2i}\sum_{n=1}^{\infty}{(-1)^{n-1}\,n\,(e^{-x})^{n}} \, dx \:\: , \:\: i\geq1.
    \end{equation}

    Summation and integration may now be interchanged. The justification is the Fubini--Tonelli theorem \cite{folland_real_1999} applied to the absolute values of the terms: since $\int_{0}^{\infty}x^{2i}e^{-nx}\,dx=(2i)!\,n^{-(2i+1)}$, the majorant series is $\sum_{n=1}^{\infty}n\,(2i)!\,n^{-(2i+1)}=(2i)!\,\sum_{n=1}^{\infty}n^{-2i}$, which converges for every $i\geq1$. Note that the series $\sum_{n\geq1}(-1)^{n-1}n\,e^{-nx}$ is not uniformly convergent on $[0,\infty)$ (its terms do not tend to zero at $x=0$) so the interchange must be justified through this majorant rather than through uniform convergence. For $i=0$ the majorant degenerates into the harmonic series and diverges, which is precisely why the case $M_0$ was evaluated separately above.

    \begin{equation}
        M_{2i} = 2 \sum_{n=1}^\infty { (-1)^{n-1}n \, \int_{0}^{\infty} x^{2i}e^{-nx} \, dx } \:\: , \:\: i\geq1.
    \end{equation}

    By making this, the integral has the shape of a Gamma Function. To be clear, let's make the transformation $u=nx$ and develop.

    \begin{equation}
        \int_{0}^{\infty} x^{2i}e^{-nx} \, dx = \int_{0}^{\infty} \left(\frac{u}{n}\right)^{2i}e^{-u} \, \left(\frac{du}{n}\right) = n^{-(2i+1)} \int_{0}^{\infty} u^{2i}e^{-u} \, du.
    \end{equation}

    By just using the definition of the Gamma function we can get the result $n^{-(2i+1)}\Gamma(2i+1)=n^{-(2i+1)}\cdot(2i)!$. We substitute this on $M_{2i}$ and then

    \begin{equation}
        M_{2i} = 2(2i)! \, \sum_{n=1}^{\infty}{ \frac{(-1)^{n-1}}{n^{2i}} } \:\: , \:\: i\geq1.
    \end{equation}

    The remaining sum is a Dirichlet eta function. The Dirichlet's eta function is defined as $\eta(s)=\sum_{n=1}^{\infty}\frac{(-1)^{n-1}}{n^s}$ and satisfies $\eta(s)=\left(1-2^{1-s}\right)\zeta(s)$ \cite{AbramowitzStegun1972Handbook}. Therefore

    \begin{equation} \label{M_generalsolution}
        M_{2i} = 2(2i)!\,(1-2^{1-2i})\zeta(2i) \:\: , \:\: i\geq1.
    \end{equation}

    It is worth noting that Eq.~(\ref{M_generalsolution}), read together with the special value $\zeta(0)=-1/2$, formally reproduces the trivial case as well, since $2\,(0)!\,(1-2)\zeta(0)=1=M_0$; the closed form is thus valid for all $i\geq0$. The first values of this function are

    \begin{table}[h!]
        \centering
        \begin{tabular}{c|c}
        $M_0$ & $1$ \\
        $M_2$ & $\pi^2/3$ \\
        $M_4$ & $7\pi^4/15$ \\
        $M_6$ & $31\pi^6/21$ 
        \end{tabular}
        \caption{First values of $M_{2i}$}
        \label{tab:M_values}
    \end{table}

    These are precisely the coefficients that appear in the Sommerfeld expansion of Fermi-liquid thermodynamics \cite{AshcroftMermin1976}; we have verified each entry of Table~\ref{tab:M_values}, and the general formula (\ref{M_generalsolution}) up to $i=3$, by direct numerical integration to 30-digit precision by making a program using \texttt{Fortran} that solves equation (\ref{M_function}) numerically for different values of $i$ and also computes the exact values that we can obtain using (\ref{M_generalsolution}). The obtained results for $1\leq i \leq 10$ are in the next table.

    \begin{table}[h!]
        \centering
        \scriptsize{
        \begin{tabular}{cccc}
        \hline 
        $i$ & Numerical $M_{2i}$ & Analytic $M_{2i}$ & Porcentual error \\
        \hline
        1  & $3.289868E+00$ & $3.289868E+00$ & $8.196E-32\,\%$ \\
        2  & $4.545757E+01$ & $4.545757E+01$ & $1.356E-32\,\%$ \\
        3  & $1.419193E+03$ & $1.419193E+03$ & $2.779E-32\,\%$ \\
        4  & $8.033622E+04$ & $8.033622E+04$ & $3.142E-32\,\%$ \\
        5  & $7.250629E+06$ & $7.250629E+06$ & $1.448E-31\,\%$ \\
        6  & $9.577710E+08$ & $9.577710E+08$ & $1.080E-32\,\%$ \\
        7  & $1.743459E+11$ & $1.743459E+11$ & $2.733E-31\,\%$ \\
        8  & $4.184494E+13$ & $4.184494E+13$ & $6.963E-31\,\%$ \\
        9  & $1.280469E+16$ & $1.280469E+16$ & $2.980E-31\,\%$ \\
        10 & $4.865799E+18$ & $4.865799E+18$ & $5.476E-32\,\%$ \\
        \hline 
        \end{tabular}}
        \caption{Numerical vs analytic $M_{2i}$ values}
        \label{tab:M_comparison}
    \end{table}

    So, the value of the conductivity is

    \begin{equation} \label{conductivity}
        \sigma = 2\sum_{i=0}^{\infty}{ (k_BT)^{2i} G^{(2i)}(\mu)(1-2^{1-2i})\zeta(2i) }.
    \end{equation}

    \subsection{The all-orders zeta-function Mott series}

    Now let's calculate $S\sigma$ in a similar way. The product ${S\sigma}$ can be calculated \cite{PhysRev.181.1336} as

    \begin{equation}
        S\sigma = -\frac{k_B}{e}\,\frac{1}{(k_BT)^2}\int_{-\infty}^{\infty}(E-\mu)G(E)f(1-f)\,dE.
    \end{equation}

    The physical content of this expression is the Cutler--Mott structure of the thermoelectric response. The same spectral kernel that determines $\sigma$ determines $S\sigma$, but weighted by one extra power of $(E-\mu)$, which measures the particle--hole asymmetry of transport about the chemical potential \cite{PhysRev.181.1336}. If $G(E)$ were exactly symmetric about $\mu$, the thermopower would vanish identically. Following the same procedure we made for the conductivity, we make a Taylor expansion of $G(E)$.

    \begin{equation}
        S\sigma = -\frac{k_B}{e}\,\frac{1}{(k_BT)^2} \sum_{n=0}^{\infty}{ \frac{G^{(n)}(\mu)}{n!} \int_{-\infty}^{\infty} (E-\mu)^{n+1}f(1-f) \, dE }.
    \end{equation}

    Because $f(1-f)$ is even, the odd $(E-\mu)^{n+1}$ terms will vanish. Therefore $n$ must be odd and we let $n=2j+1$ and then 

    \begin{equation}
        S\sigma = -\frac{k_B}{e}\,\frac{1}{(k_BT)^2} \sum_{j=0}^{\infty}{ \frac{G^{(2j+1)}(\mu)}{(2j+1)!} \int_{-\infty}^{\infty} (E-\mu)^{2j+2}f(1-f) \, dE }.
    \end{equation}

    This integral is similar to the previous that we solve, by making the same transformation we can see that

    \begin{equation}
        \int_{-\infty}^{\infty} (E-\mu)^{2j+2}f(1-f) \, dE = (k_BT)^{2j+3}\cdot2(2j+2)!\,(1-2^{-2j-1})\zeta(2j+2).
    \end{equation}

    Then

    \begin{equation} \label{pre-seebeck}
        S\sigma = -2\frac{k_B}{e}\sum_{j=0}^{\infty}{ G^{(2j+1)}(\mu)(2j+2)(k_BT)^{2j+1}(1-2^{-2j-1})\zeta(2j+2) }.
    \end{equation}

    Note that only the \emph{odd} derivatives of $G$ enter Eq.~(\ref{pre-seebeck}), while only the even derivatives enter the conductivity (\ref{conductivity}). The thermopower is generated exclusively by the antisymmetric part of the transport function around $\mu$, in agreement with the qualitative particle--hole argument above. With this result we can obtain a more general Mott-Formula. By replacing (\ref{conductivity}) in (\ref{pre-seebeck}) we get the all-orders \textit{zeta-function Mott series}

    \begin{equation} \label{Generalized_Seebeck}
        S_{\mathrm{GEN}} = -\frac{k_B^2T}{e}\frac{\sum_{i=0}^{\infty}{(2i+2)(1-2^{-2i-1})(k_BT)^{2i}G^{(2i+1)}(\mu)\zeta(2i+2)}}{\sum_{i=0}^{\infty}{(1-2^{1-2i})(k_BT)^{2i}G^{(2i)}(\mu)\zeta(2i)}}.
    \end{equation}

    We can also write this equation in terms of $M_m$ as

    \begin{equation}
        S_{\mathrm{GEN}}=-\frac{k_B^2T}{e}\,\frac{\displaystyle\sum_{i=0}^{\infty}(k_BT)^{2i}\,\frac{G^{(2i+1)}(\mu)}{(2i+1)!}\,M_{2i+2}}{\displaystyle\sum_{i=0}^{\infty}(k_BT)^{2i}\,\frac{G^{(2i)}(\mu)}{(2i)!}\,M_{2i}}.
    \end{equation}

    Equation~(\ref{Generalized_Seebeck}) resums (in closed indexed form) every order of the Sommerfeld expansion of the Seebeck coefficient. We emphasize its scope of validity, which it inherits from the Cutler--Mott framework \cite{PhysRev.181.1336}. It holds for independent carriers in the relaxation-time (or static-disorder) picture, with a transport function $G(E)$ that is analytic in a neighborhood of $\mu$ wide enough for the expansion to represent it; strong inelastic scattering, interaction-driven hydrodynamic regimes such as the Dirac fluid of graphene near charge neutrality \cite{crossno_observation_2016}, and phonon-drag contributions lie outside its scope. For polynomial $G(E)$ the two series in (\ref{Generalized_Seebeck}) terminate, and the resummation is exact and non-perturbative in $k_BT/\mu$.

    \subsection{Recovery of the standard Mott formula}

    By keeping only the term with $i=0$ in both series of Eq.~(\ref{Generalized_Seebeck}), we obtain

    \begin{equation}
        S = -\frac{k_B^2T}{e}\,\frac{2(1-2^{-1})\zeta(2)}{(1-2)\zeta(0)}\,\frac{G^{\prime}(\mu)}{G(\mu)}.
    \end{equation}

    Then, evaluating the zeta function and simplifying the expression we get

    \begin{equation}
        S = -\frac{\pi^2k_B^2T}{3e}\,\frac{d\,\ln{G(E)}}{dE}\big{|}_{E=\mu}.
    \end{equation}

    Here we used $\zeta(2)=\pi^2/6$ and $\zeta(0)=-1/2$, so that the numerical prefactor becomes $(\pi^2/6)/(1/2)=\pi^2/3$. The standard Mott-Formula is an approximation for low temperatures, in this limit we can see that $\mu\longrightarrow E_F$ where $E_F$ is the Fermi Energy. Also, keeping only $i=0$ in Eq.~(\ref{conductivity}) and using $2(1-2)\zeta(0)=1$, we get

    \begin{equation}
        \sigma = G(E_F).
    \end{equation}

    Therefore, we recover the conventional Mott-Formula

    \begin{equation}
        S = -\frac{\pi^2k_B^2T}{3e}\,\frac{d\,\ln{\sigma(E)}}{dE}\big{|}_{E=E_F}.
    \end{equation}

    This consistency check confirms that Eq.~(\ref{Generalized_Seebeck}) contains the textbook result \cite{ModernCondensedMatterPhysics,goldsmid_introduction_2016} as its leading term, with all higher terms supplying corrections of relative order $(k_BT/\mu)^{2}$ and beyond. In the degenerate regime $k_BT\ll\mu$ these corrections are small; in the intermediate regime relevant to room-temperature graphene devices, where $E_F$ may be only a few times $k_BT$, they are not, and the full series must be used.

\section{All orders Zeta-Function Expansion of the Graphene PTE Transport Integrals}

    The photothermoelectric (PTE) effect is one of the dominant mechanisms of photoresponse in graphene devices: absorbed light heats the electronic system well above the lattice temperature, and the resulting hot-carrier temperature gradient drives a photovoltage through the Seebeck effect \cite{xu_photothermoelectric_2010,gabor_hot_2011}. The efficiency of this process is controlled by how slowly the hot carriers cool---in disordered graphene, predominantly through disorder-assisted (supercollision) electron--phonon scattering \cite{song_disorder_2012}---and by the ultrafast thermalization dynamics of the photoexcited carriers \cite{tomadin_ultrafast_2018}. PTE detectors based on graphene now form an important branch of photothermoelectric photodetector technology \cite{lu_progress_2019}. The theoretical description of these devices rests on the electronic transport coefficients $\sigma$, $S$ and $\kappa_e$ (the electronic thermal conductivity), which in the linear-response regime are all generated by a small set of transport integrals. In this section we show that, for the standard conductivity model of graphene \cite{antidormi_optimizing_2021}, the zeta-function machinery of Section 1 evaluates these integrals exactly.

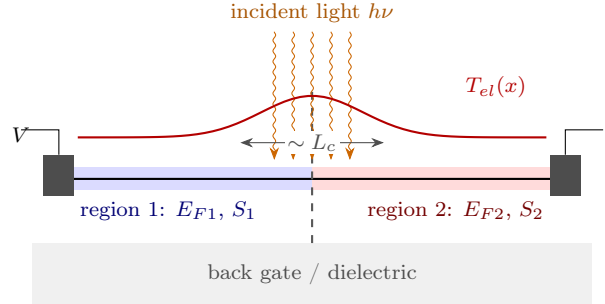
\begin{figure}[h!]
        \centering
        \begin{tikzpicture}[>=Stealth, font=\small]
            \begin{scope}
                \fill[gray!12] (-0.2,-1.85) rectangle (7.2,-1.05);
                \node[gray!60!black, font=\scriptsize] at (3.5,-1.45) {back gate / dielectric};
                \fill[blue!14] (0.2,-0.35) rectangle (3.5,-0.05);
                \fill[red!14]  (3.5,-0.35) rectangle (6.8,-0.05);
                \draw[thick] (0.2,-0.2) -- (6.8,-0.2);
                \draw[dashed, thick, gray!70!black] (3.5,-1.05) -- (3.5,1.05);
                \node[font=\scriptsize, blue!45!black] at (1.6,-0.62) {region 1: $E_{F1}$, $S_1$};
                \node[font=\scriptsize, red!45!black]  at (5.4,-0.62) {region 2: $E_{F2}$, $S_2$};
                \fill[black!70] (-0.05,-0.42) rectangle (0.35,0.12);
                \fill[black!70] (6.65,-0.42) rectangle (7.05,0.12);
                \node[font=\scriptsize] at (-0.35,0.38) {$V$};
                \draw (0.15,0.12) -- (0.15,0.45) -- (-0.35,0.45);
                \draw (6.85,0.12) -- (6.85,0.45) -- (7.35,0.45);
                \foreach \s in {-0.5,-0.25,0,0.25,0.5}{
                    \draw[->, orange!80!black, decorate,
                          decoration={snake, amplitude=.5pt, segment length=4pt}]
                          (3.5+\s,1.75) -- (3.5+\s,0.05);
                }
                \node[orange!65!black, font=\scriptsize] at (3.5,2.0) {incident light $h\nu$};
                \draw[red!70!black, thick]
                    plot[domain=0.4:6.6, samples=60]
                    (\x, {0.35+0.55*exp(-(\x-3.5)*(\x-3.5)/0.9)});
                \node[red!70!black, font=\scriptsize, anchor=west] at (5.4,1.02) {$T_{el}(x)$};
                \draw[<->, gray!60!black] (2.55,0.28) -- (4.45,0.28);
                \node[gray!60!black, font=\scriptsize, fill=white, inner sep=1pt] at (3.5,0.28) {$\sim L_c$};
            \end{scope}
        \end{tikzpicture}
        \caption{Schematic of the problem addressed. A graphene photothermoelectric junction. Light absorbed near the interface raises the electronic temperature $T_{el}$ above the lattice temperature over a region of size set by the cooling length $L_c$; because the two sides are gated to different Fermi energies, their Seebeck coefficients differ and a photovoltage $V_{\mathrm{PTE}}=(S_2-S_1)\Delta T$ appears between the contacts.}
        \label{fig:pte_schematic}
    \end{figure}

    Three features of graphene PTE devices make them the setting in which an all-orders treatment stops being a formal refinement and becomes necessary. First, the operating point sits in the wrong regime for the Mott formula. A graphene photodetector is gated to a few hundred meV at most, and at room temperature $k_BT\simeq26$~meV, so $E_F/k_BT$ is typically of order a few---exactly the intermediate-degeneracy window in which the leading-order expansion is least reliable, and, as Section~\ref{subsec:ZTe} shows, exactly where the thermoelectric response is largest (the small parameter of the Sommerfeld expansion is not small in the device). Second, the response is read out through a difference of Seebeck coefficients. The photovoltage generated at a junction between two regions with different doping is $V_{\mathrm{PTE}}=(S_2-S_1)\,\Delta T$, so a systematic overestimation of $S$ propagates directly into the predicted responsivity, and the error does not cancel between the two sides of the junction unless both are treated consistently. Third, the same three moments $\mathcal{K}_0$, $\mathcal{K}_1$ and $\mathcal{K}_2$ that fix the electrical conductivity $\sigma$ and the Seebeck coefficient $S$ also fix the electronic thermal conductivity $\kappa_e$, and through it the cooling length $L_c\propto\sqrt{\kappa_e}$ that sets the spatial extent of the photoactive region. Errors in the transport integrals therefore contaminate not only the magnitude of the signal but also the effective device area that produces it, and hence the noise-equivalent power. A closed-form, temperature-exact evaluation of $\mathcal{K}_j$ fixes all three at once. Figure~\ref{fig:pte_schematic} summarises the geometry and the chain of quantities involved, as in the Ref.\cite{antidormi_optimizing_2021}.

    \subsection{Transport integrals and photothermoelectric observables}

    In their paper, Antidormi \& Cummings \cite{antidormi_optimizing_2021} define the transport integrals

    \begin{equation} \label{Graphene_transport_integrals}
        \mathcal{K}_j = \int_{-\infty}^{\infty} (E-E_F)^j \, \tilde{\sigma}(E) \, \left(-\frac{\partial f}{\partial E}\right) dE \:\: ; \:\: j=0,1,2.
    \end{equation}

    Where $\tilde\sigma$ is the dimensionless conductivity defined by $\tilde\sigma=\sigma/(2e^2/h)$, $E_F$ is the Fermi energy, $f$ is the Fermi-Dirac Function, $e$ is the electrical charge, $h$ is Planck's constant and $\mathcal{K}_j$ are the transport moments. These are the standard moments of linear-response (Onsager) transport theory \cite{sivan_imry_1986,guttman_thermopower_1995,guttman_thermal_1995}, written in the Landauer-type normalization natural for two-dimensional conductors: $\mathcal{K}_0$ fixes the electrical conductance, $\mathcal{K}_1$ measures the particle--hole asymmetry responsible for the thermopower, and the combination $\mathcal{K}_2-\mathcal{K}_1^2/\mathcal{K}_0$ fixes the electronic heat conductance at zero electrical current. They are, up to the normalization $2e^2/h$ and the identification of the mathematical role $G(E)\leftrightarrow\sigma(E)$, the same moment integrals that appeared in our derivation of the zeta-function Mott series in Section 1. This identification is the bridge that lets us apply the all-orders expansion directly to the graphene PTE problem. Using the definition of $\mathcal{K}_j$, Sivan and Imry \cite{sivan_imry_1986} define

    \begin{subequations} \label{Coeficients}
        \begin{align}
            S &= -\frac{1}{eT} \frac{\mathcal{K}_1}{\mathcal{K}_0}, \\
            \kappa_e &= \frac{2}{hT} \left( \mathcal{K}_2-\frac{\mathcal{K}_1^2}{\mathcal{K}_0} \right), \\
            L_c &= \sqrt{\frac{\kappa_e}{\gamma\,C_{el}}}.
        \end{align}
    \end{subequations}

    Where $S$ is the Seebeck Coefficient, $\kappa_e$ is the electronic thermal conductivity, $L_c$ is the cooling length, $\gamma$ is the cooling ratio, and $C_{el}$ is the specific heat. These quantities are useful to define the thermal voltage $V_{\mathrm{PTE}}$, the thermal current $I_{\mathrm{PTE}}$ and the noise-equivalent power $\mathrm{NEP}$. The cooling length $L_c$ deserves a comment: it is the distance over which the hot electronic system equilibrates with the lattice, and it enters the device response because only junction regions within $\sim L_c$ of the illuminated spot contribute to the photovoltage \cite{antidormi_optimizing_2021,song_disorder_2012}. We can rewrite (\ref{Graphene_transport_integrals}) as

    \begin{equation}
        \mathcal{K}_j = \frac{1}{k_BT} \int_{-\infty}^{\infty} (E-E_F)^j \, \tilde{\sigma}(E) \, f(1-f) dE \:\: ; \:\: j=0,1,2.
    \end{equation}

    \subsection{All-orders expansion in zeta functions}

    This integral has the same structure as the transport integrals that we previously solved, so it is convenient to solve it the same way. If we expand $\tilde\sigma$ around $E_F$ we get

    \begin{equation}
        \mathcal{K}_j = \frac{1}{k_BT} \sum_{n=0}^{\infty}\frac{\tilde{\sigma}^{(n)}(E_F)}{n!} \int_{-\infty}^{\infty} (E-E_F)^{j+n} \, f(1-f) dE \:\: ; \:\: j=0,1,2.
    \end{equation}

    Then

    \begin{equation}
        \mathcal{K}_j = \sum_{n=0}^{\infty}\frac{\tilde{\sigma}^{(n)}(E_F)}{n!} \, (k_BT)^{j+n} \, M_{j+n} \:\: ; \:\: j=0,1,2.
    \end{equation}

    The function $M_{2i}$ is defined in (\ref{M_function}), and it's explicit form is in (\ref{M_generalsolution}), let's rewrite it here.

    \begin{equation*}
        M_m = \int_{-\infty}^{\infty} x^m\frac{e^x}{(1+e^x)^2}\,dx \:\: ; \:\: x = \frac{E-E_F}{k_BT}.
    \end{equation*}

    Also, in order to have non-zero values $m$ must be even. So the remaining terms are

    \begin{equation*}
        M_m = 2\,m!\,(1-2^{1-m})\zeta(m) \:\: ; \:\: m \:\: \text{even},\:\: m\geq2.
    \end{equation*}

    Therefore $j+n$ must be even. Also, it is important to notice that $M_0=1$ and $M_{2i+1}=0$. With all of this, we can see that for $j=0$ only $n=2i$ contribute, for $j=1$ only $n=2i+1$ contribute and for $j=2$ only $n=2i$ contribute, so we get

    \begin{equation} \label{Graphene_transport_series}
        \begin{split}
            \mathcal{K}_0 &= \sum_{i=0}^{\infty} \frac{\tilde{\sigma}^{(2i)}(E_F)}{(2i)!} \, (k_BT)^{2i} \, M_{2i}, \\
            \mathcal{K}_1 &= \sum_{i=0}^{\infty} \frac{\tilde{\sigma}^{(2i+1)}(E_F)}{(2i+1)!} \, (k_BT)^{2i+2} \, M_{2i+2}, \\
            \mathcal{K}_2 &= \sum_{i=0}^{\infty} \frac{\tilde{\sigma}^{(2i)}(E_F)}{(2i)!} \, (k_BT)^{2i+2} \, M_{2i+2}.
        \end{split}
    \end{equation}

    Equation~(\ref{Graphene_transport_series}) makes the parity structure of the transport coefficients explicit. $\mathcal{K}_0$ and $\mathcal{K}_2$ probe the even derivatives of $\tilde\sigma$ at the Fermi level, while $\mathcal{K}_1$---and hence the Seebeck coefficient---probes only the odd ones, exactly as found for the general transport function in Section 1.

    \subsection{Exact termination for the quadratic conductivity model}

    To continue solving this problem, we need an explicit shape for $\tilde\sigma$. Antidormi and Cummings model the Graphene by

    \begin{equation} \label{graphene_model}
        \sigma(E) = \sigma_{\min} + \mu_ce\,n_c(E).
    \end{equation}

    Where $\mu_c$ is the carrier mobility and $n_c(E)$ is the carrier density, given by\footnote{Antidormi and Cummings denote the carrier mobility by $\mu$; here we write $\mu_c$ to avoid any confusion with the chemical potential $\mu$ of Section 1. Throughout the present section the conductivity is evaluated at fixed $E_F$, so that $\mu\to E_F$ is understood.}

    \begin{equation}
        n_c(E) = \frac{E^2}{\pi\hbar^2v_F^2}.
    \end{equation}

    Here $v_F$ is the Fermi velocity and $\hbar$ is the reduced Planck constant $\hbar=h/2\pi$. This phenomenological form captures the two essential features of transport in real graphene samples \cite{castroneto_electronic_2009,dassarma_electronic_2011}.
    
    Graphene conductivity models more generally can be organized by the energy dependence of the relaxation time, $\tau(E)\propto|E|^m$, with short-range, unscreened-Coulomb, and screened-Coulomb impurity scattering each corresponding to a different exponent $m$, as catalogued for monolayer graphene within a Boltzmann framework by Ray \& Sarkar \cite{ray_mott_2025}. The conductivity model of Eq.~(\ref{graphene_model}) (a constant offset plus a quadratic term) is not itself a member of that single-exponent monomial family, but sits within the same general class of energy-dependent conductivity models used to describe graphene transport; we work exclusively with this quadratic-plus-offset form, for which, as shown below, the zeta-function series of Section 1 terminates exactly. Away from the Dirac point the conductivity grows linearly with the carrier density (i.e., quadratically with energy, since $n_c\propto E^2$ for the linear Dirac dispersion), while at the charge-neutrality point it saturates at a residual value $\sigma_{\min}$ produced by electron--hole puddles and disorder. The same quadratic-in-$E_F$ scaling of $\sigma$ underlies the standard theory of the graphene thermopower \cite{hwang_thermopower_2009} and is consistent with the measurements of Refs.~\cite{zuev_thermoelectric_2009,wei_anomalous_2009}. By making $c=\mu_ce/\pi\hbar^2v_F^2$ we can rewrite (\ref{graphene_model}) as

    \begin{equation}
        \sigma(E) = \sigma_{\min} + cE^2.
    \end{equation}

    Because this is a quadratic equation, $\sigma^{(k)}=0$ for $k\geq3$ and the series given in (\ref{Graphene_transport_series}) terminates exactly as following

    \begin{equation}
        \begin{split}
            \mathcal{K}_0 &= \frac{\tilde{\sigma}(E_F)}{0!}M_0 + \frac{\tilde{\sigma}^{\prime\prime}(E_F)}{2!}\,(k_BT)^2M_2, \\
            \mathcal{K}_1 &= \frac{\tilde{\sigma}^{\prime}(E_F)}{1!}\,(k_BT)^2M_2, \\
            \mathcal{K}_2 &= \frac{\tilde{\sigma}(E_F)}{0!}(k_BT)^2M_2 + \frac{\tilde{\sigma}^{\prime\prime}(E_F)}{2!}\,(k_BT)^4M_4.
        \end{split}
    \end{equation}

    We stress that this termination is an algebraic identity of the quadratic model, valid at any temperature and Fermi energy for which the model conductivity itself applies. All the results below are therefore exact within the model of Eq.~(\ref{graphene_model}), and their scope of validity is that of the model: a Fermi-liquid description of graphene away from the hydrodynamic regime near charge neutrality \cite{crossno_observation_2016}, with the conductivity evaluated at fixed $E_F$. \\
    
    Defining $\tilde{c}\equiv c/(2e^2/h)=ch/2e^2$, we can rewrite $\tilde{\sigma}$ as $\tilde{\sigma}=\tilde{c}(\sigma_{\min}/c\:+\:E^2)$, so the derivatives are $\tilde{\sigma}^{\prime}=2\tilde{c}E$ and $\tilde{\sigma}^{\prime\prime}=2\tilde{c}$. The needed values of $M_{2i}$ were computed before and they are in the table (\ref{tab:M_values}). Replacing all of this in the finite series for $\mathcal{K}_j$ we obtain

    \begin{equation}
        \begin{split}
            \mathcal{K}_0 &= \tilde{c}\left[ \frac{\sigma_{\min}}{c} + E_F^2 + \frac{\pi^2}{3}(k_BT)^2 \right], \\
            \mathcal{K}_1 &= 2\tilde{c}E_F(k_BT)^2\frac{\pi^2}{3}, \\
            \mathcal{K}_2 &= \tilde{c}\left[ \left(\frac{\sigma_{\min}}{c}+E_F^2\right)\frac{\pi^2}{3}(k_BT)^2 + \frac{7\pi^4}{15}(k_BT)^4 \right]. 
        \end{split}
    \end{equation}

    It is convenient to rewrite this to a more simple shape. In order to do this, let's factorize $E_F^{j+2}$ in each situation.

    \begin{equation}
        \begin{split}
            \mathcal{K}_0 &= \tilde{c}E_F^2 \left[ \frac{\sigma_{\min}}{cE_F^2} + 1 + \frac{\pi^2}{3}\frac{(k_BT)^2}{E_F^2} \right], \\
            \mathcal{K}_1 &= \tilde{c}E_F^3 \left[ \frac{2\pi^2}{3}\frac{(k_BT)^2}{E_F^2} \right], \\
            \mathcal{K}_2 &= \tilde{c}E_F^4 \left[ \frac{(k_BT)^2}{E_F^2}\left(\frac{\sigma_{\min}}{cE_F^2}+1\right)\frac{\pi^2}{3} + \frac{7\pi^4}{15}\frac{(k_BT)^4}{E_F^4} \right]. 
        \end{split}
    \end{equation}

    Now we define two parameters that will make the following analysis more simple.

    \begin{equation} \label{Graphene_transport_abreviations}
        t\equiv\frac{k_BT}{E_F} \:\:\:\: , \:\:\:\: l\equiv\frac{\sigma_{\min}}{cE_F^2}.
    \end{equation}

    Both parameters are dimensionless and have a different physical meaning. $t$ measures the degeneracy of the electron gas (the degenerate Fermi-liquid regime corresponds to $t\ll1$), while $l$ measures the relative weight of the residual disorder-limited conductivity $\sigma_{\min}$ against the gate-induced contribution $cE_F^2$; a high-quality, strongly doped sample has $l\ll1$. By making this we are left with

    \begin{equation} \label{Graphene_transport_numbers}
        \begin{split}
            \mathcal{K}_0 &= \tilde{c}E_F^2 \left[ l + 1 + \frac{\pi^2}{3}t^2 \right], \\
            \mathcal{K}_1 &= \tilde{c}E_F^3 \left[ \frac{2\pi^2}{3}t^2 \right], \\
            \mathcal{K}_2 &= \tilde{c}E_F^4 \left[ t^2(l+1)\frac{\pi^2}{3} + \frac{7\pi^4}{15}t^4 \right]. \\
        \end{split}
    \end{equation}

    \subsection{Closed-form Seebeck coefficient}

    These equations are exact, and a direct consequence is the closed form of (\ref{Coeficients}). We can do a few things with this results, so let's start by calculating the Seebeck coefficient. By just replacing (\ref{Graphene_transport_numbers}) we get

    \begin{equation*}
        S = -\frac{1}{eT}\frac{\tilde{c}E_F^3 \left[ \frac{2\pi^2}{3}t^2 \right]}{\tilde{c}E_F^2 \left[ l + 1 + \frac{\pi^2}{3}t^2 \right]} = -\frac{E_F\,t}{eT} \frac{2\pi^2 t/3}{l+1+\pi^2t^2/3}. 
    \end{equation*}

    It is easy to notice that $E_F\,t/T=k_B$, so $E_F\,t/(eT)=k_B/e$, therefore the Seebeck coefficient is

    \begin{equation} \label{Seebeck_graphene}
        S = -\frac{k_B}{e} \frac{2\pi^2 t/3}{l+1+\pi^2t^2/3}.
    \end{equation}

    Equation~(\ref{Seebeck_graphene}) has the correct limiting behaviors. For $t\ll1$ it reduces to the Mott result for the quadratic model, $S\to-(k_B/e)\,2\pi^2t/[3(l+1)]$, linear in temperature as expected for a degenerate conductor \cite{AshcroftMermin1976,hwang_thermopower_2009}. As $t$ grows, however, the term $\pi^2t^2/3$ in the denominator (please notice that this term is absent from the Mott formula) saturates the growth of $|S|$, which reaches the maximum value $|S|_{\max}=(k_B/e)\,\pi/\sqrt{3(l+1)}$ at $t^2=3(l+1)/\pi^2$ and then decreases (both statements follow by elementary differentiation of Eq.~(\ref{Seebeck_graphene})). This built-in saturation is precisely the mechanism by which the exact result cures the severe overestimation of $S$ produced by the bare Mott formula in high-mobility graphene, an overestimation reported numerically in Ref.~\cite{antidormi_optimizing_2021}; here it is obtained in closed form.

    In order to have a better understanding of this, we can compute $S$ for different values of $l$.

    \begin{figure}[h!]
        \centering
        \includegraphics[width=0.75\linewidth]{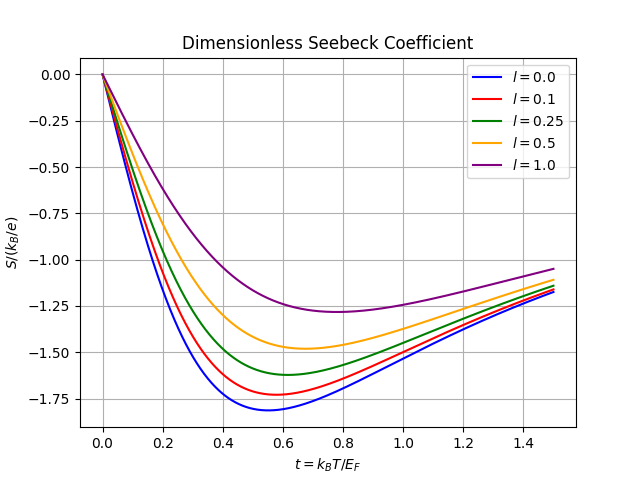}
        \caption{Dimensionless Seebeck coefficient $S/(k_B/e)$ of Eq. (\ref{Seebeck_graphene}) as a function of $t = k_BT/E_F$, for several values of the disorder parameter $l=\sigma_{\min}/(cE_F^2)$. Every curve starts at zero, reaches a single extremum (most negative value) at $t_{\mathrm{opt}} = \sqrt{3(l+1)}/\pi$, and decays back toward zero. The clean limit $l=0$ gives the largest attainable $|S|$, $|S|_{\max} = \pi/\sqrt{3} \approx 1.8138$, at $t_{\mathrm{opt}} \approx 0.5513$, i.e.\ $E_F \simeq 1.8138\,k_BT$. All curves approach the common, $l$-independent asymptote $S \to -2(k_B/e)/t$ at large $t$.}
        \label{fig:seebeck}
    \end{figure}

    Figure~\ref{fig:seebeck} shows the Seebeck coefficient in units of $k_B/e$. Every curve starts at $S=0$. As $t \to 0$, the thermal window collapses onto $E_F$ and, to leading order, $S$ reduces to the ordinary Mott result for the quadratic model,

    \begin{equation}
        \left.\frac{dS}{dt}\right|_{t=0} = -\frac{2\pi^2}{3(l+1)}\,\frac{k_B}{e}.
    \end{equation}

    We can notice that the $\pi^2t^2/3$ term dominates the denominator of Eq. (\ref{Seebeck_graphene}) and every curve collapses onto the same, $l$-independent asymptote $S \to -2(k_B/e)/t$, visible in Fig.~\ref{fig:seebeck} as the merging of all five curves at large $t$. Between these two limits each curve passes through a single extremum, located at
    
    \begin{equation}
        t_{\mathrm{opt}} = \frac{\sqrt{3(l+1)}}{\pi}, \qquad S_{\mathrm{extreme}} = -\frac{\pi}{\sqrt{3(l+1)}}\,\frac{k_B}{e},
        \label{eq:seebeck_extremum}
    \end{equation}

    obtained directly from $dS/dt=0$ in Eq. (\ref{Seebeck_graphene}). Table~\ref{tab:seebeck} collects the optima for the values of $l$ plotted; the entries were produced with a program made in \texttt{Fortran} and agree with Eq.~\eqref{eq:seebeck_extremum} to all digits shown, as well as with an independent golden-section search carried out by the same program as a runtime cross-check.

    \begin{table}[h!]
        \centering
        \begin{tabular}{ccccc}
        \hline
            $l$ & $t_{\mathrm{opt}}$ & $E_F^{\mathrm{opt}}/k_BT$ & $S_{\mathrm{extreme}}$ & $S(t=1.5)$ \\
            \hline
            0    & 0.551329 & 1.813799 & $-1.813799$ & $-1.174645$ \\
            0.1  & 0.578239 & 1.729390 & $-1.729390$ & $-1.160829$ \\
            0.25 & 0.616404 & 1.622311 & $-1.622311$ & $-1.140704$ \\
            0.5  & 0.675237 & 1.480961 & $-1.480961$ & $-1.108670$ \\
            1    & 0.779697 & 1.282550 & $-1.282550$ & $-1.049712$ \\
            \hline
    \end{tabular}
    \caption{Position and depth of the extremum of $S/(k_B/e)$, from Eq.~\eqref{eq:seebeck_extremum}, for the curves of Fig.~\ref{fig:seebeck}. The last column gives the endpoint value at $t=1.5$ for reference. The clean-limit entry $S_{\mathrm{extreme}} = -1.813799$ at $t_{\mathrm{opt}} = 0.551329$ is the universal ceiling of $|S|$ within the model.}
    \label{tab:seebeck}
    \end{table}

    Two features are worth noting. First, comparing Eq.~\eqref{eq:seebeck_extremum} directly to $t_{\mathrm{opt}}$ shows the extremal value satisfies the exact identity $S_{\mathrm{extreme}}\, t_{\mathrm{opt}} = -k_B/e$, i.e.\ $|S_{\mathrm{extreme}}|$ equals, in these units, the optimal ratio $E_F^{\mathrm{opt}}/k_BT$ itself (compare columns 2--4 of Table~\ref{tab:seebeck}). This relation is a straightforward corollary of Eq.~\eqref{eq:seebeck_extremum}. 
    An identity of the same structure has a precedent in a different transport setting: for resonant scattering in a one-dimensional disordered Landauer--B\"uttiker conductor, M\"uller, Smit \& Sigrist find $|T_{\max}\cdot S_{\max}| = |\Delta E|/2e$, independent of the number of impurities $N$ \cite{muller_resonant_2015}. Our identity is the diffusive-transport analogue of their Eq.~(37); the two are derived for different models (a terminating Sommerfeld series here versus a single resonant conductance peak there) and involve different energy scales ($k_BT$ and the polynomial structure of $G(E)$ here, the resonance detuning $\Delta E$ there), but we note the precedent explicitly rather than presenting the recurrence of an energy-scale-independent extremal product as unprecedented.

    Second, unlike the Lorenz ratio of Fig. \ref{fig:lorenz} -- whose curves cross one another in a narrow window near $t \approx 0.6$ -- the Seebeck curves here never cross. Since $\partial S/\partial l = \tfrac{2\pi^2 t}{3}\big[(l+1)+\pi^2t^2/3\big]^{-2} > 0$ for every $t>0$, $S$ increases monotonically (becomes less negative) with $l$ at every fixed $t$, so the curves remain strictly ordered by $l$ across the entire range shown. The residual conductivity $\sigma_{\min}$ therefore suppresses the thermopower uniformly at every temperature, in contrast to its more intricate, sign-changing effect on the heat-transport ratio $\mathcal{L}/\mathcal{L}_0$.

    \subsection{Lorenz ratio and the Wiedemann--Franz correction}

    Now let's continue with the electronic thermal conductivity. From (\ref{Graphene_transport_integrals}), with $j=0$ we get

    \begin{equation}
        \mathcal{K}_0 = \int_{-\infty}^{\infty} \tilde{\sigma}(E) \left( -\frac{\partial f}{\partial E} \right) \, dE.
    \end{equation}

    The conductivity is therefore

    \begin{equation}
        \sigma = \frac{2e^2}{h} \int_{-\infty}^{\infty} \tilde{\sigma}(E) \left( -\frac{\partial f}{\partial E} \right) \, dE = \frac{2e^2}{h}\mathcal{K}_0.
    \end{equation}

    Also, it is good to remember that the electronic thermal conductivity is also defined by the Lorenz number $\mathcal{L}$ as $\kappa_e=\mathcal{L}\sigma T$. The Wiedemann--Franz law states that in a degenerate Fermi liquid with elastic scattering $\mathcal{L}$ takes the universal Sommerfeld value $\mathcal{L}_0=(\pi^2/3)(k_B/e)^2$ \cite{AshcroftMermin1976,jonson_mahan_1980}; deviations of the ratio $\mathcal{L}/\mathcal{L}_0$ from unity quantify the breakdown of this law, which in graphene can be dramatic in the hydrodynamic regime \cite{crossno_observation_2016} but remains finite and computable in the disordered Fermi-liquid regime considered here. Replacing all of this in (\ref{Coeficients}) leaves us with

    \begin{equation}
        \frac{2}{hT}\left( \mathcal{K}_2 - \frac{\mathcal{K}_1^2}{\mathcal{K}_0} \right) = \mathcal{L}\left( \frac{2e^2}{h}\mathcal{K}_0 \right)T.
    \end{equation}

    Then

    \begin{equation}
        \mathcal{L} = \frac{\mathcal{K}_2 - \mathcal{K}_1^2/\mathcal{K}_0}{e^2T^2\mathcal{K}_0}.
    \end{equation}

    Now we divide by $\mathcal{L}_0=(\pi^2/3)(k_B/e)^2$ and we obtain

    \begin{equation}
        \frac{\mathcal{L}}{\mathcal{L}_0} = \frac{3}{\pi^2(k_BT)^2} \frac{\mathcal{K}_2 - \mathcal{K}_1^2/\mathcal{K}_0}{\mathcal{K}_0}.
    \end{equation}

    Here, we substitute (\ref{Graphene_transport_numbers}) to obtain

    \begin{equation}
        \frac{\mathcal{L}}{\mathcal{L}_0} = \frac{3}{\pi^2(k_BT)^2} \frac{\tilde{c}E_F^4 \left[ (l+1)\frac{\pi^2}{3}t^2 + \frac{7\pi^4}{15}t^4 \right] - \frac{\tilde{c}^2E_F^6 (\frac{2\pi^2}{3}t^2)^2}{\tilde{c}E_F^2 (l+1+\frac{\pi^2}{3}t^2)}}{\tilde{c}E_F^2 (l+1+\frac{\pi^2}{3}t^2)}.
    \end{equation}

    We can directly cancel some terms and rearrange the equation, by doing that we are left with

    \begin{equation}
        \frac{\mathcal{L}}{\mathcal{L}_0} = \frac{3E_F^2}{\pi^2(k_BT)^2} \frac{(l+1)\frac{\pi^2}{3}t^2 + \frac{7\pi^2}{5}t^2\frac{\pi^2}{3}t^2 - \frac{\pi^2}{3}t^2 \frac{(4\pi^2t^2/3)}{(l+1+\pi^2t^2/3)}}{(l+1) + \frac{\pi^2}{3}t^2}.
    \end{equation}

    The term $3E_F^2/[\pi^2(k_BT)^2]$ can be written as $3/(\pi^2t^2)$, so we finally get

    \begin{equation} \label{Lorenz_graphene}
        \frac{\mathcal{L}}{\mathcal{L}_0} = \frac{(l+1) + 7\pi^2t^2/5 - (4\pi^2t^2/3)/[(l+1)+\pi^2t^2/3]}{(l+1) + \pi^2t^2/3}.
    \end{equation}

    In the degenerate limit $t\to0$ this ratio tends to unity, recovering the Wiedemann--Franz law exactly, as it must \cite{jonson_mahan_1980}. At finite $t$ the deviation is governed by a competition between two thermal corrections of opposite tendency. The term $7\pi^2t^2/5$, coming from the fourth moment $M_4$ in $\mathcal{K}_2$, enhances the heat conduction per carrier, while the bipolar-like term $-(4\pi^2t^2/3)/[(l+1)+\pi^2t^2/3]$, coming from the $\mathcal{K}_1^2/\mathcal{K}_0$ subtraction, suppresses it because part of the energy current is convected by the particle current. The partial cancellation between these two contributions explains the empirical robustness of the Wiedemann--Franz law for this model at moderate temperatures, noted in Ref.~\cite{antidormi_optimizing_2021} for $E_F>2k_BT$; Eq.~(\ref{Lorenz_graphene}) quantifies the residual deviation in closed form.
    
    A different, purely empirical route to the Lorenz number is due to Kim \emph{et al.}~\cite{kim_lorenz_2015}, who propose $L=1.5+\exp(-|S|/116)$ (in units of $10^{-8}\,\mathrm{W\Omega K^{-2}}$, with $S$ in $\mu$V/K) as a fit within the single-parabolic-band, acoustic-phonon-scattering framework, where $L$ and $S$ are both parametric functions of a single reduced chemical potential alone. That result and Eq.~(\ref{Lorenz_graphene}) are complementary routes to the Lorenz number, distinct in origin (an empirical fit to a Fermi-integral transport model there, an exact closed form within the Mott framework here) rather than competing derivations of the same quantity. \\

    In order to have a better understanding of this, we can compute the ratio $\mathcal{L}/\mathcal{L}_0$ for diferent values of $l$.

    \begin{figure}[h!]
        \centering
        \includegraphics[width=0.75\textwidth]{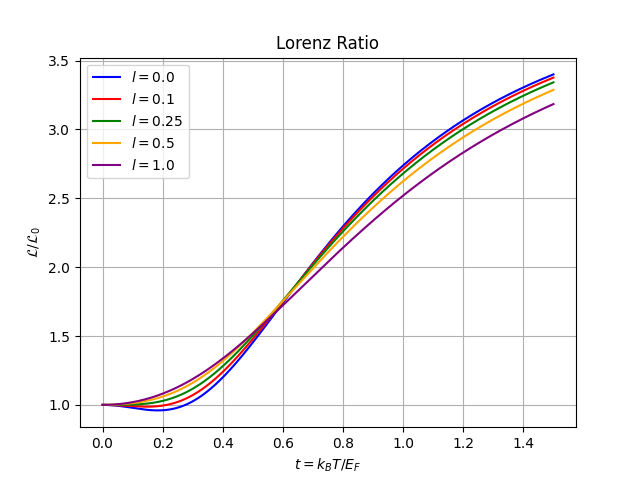}
        \caption{Lorenz ratio $\mathcal{L}/\mathcal{L}_0$ of Eq.~(\ref{Lorenz_graphene}) as a function of $t=k_BT/E_F$, for several values of the disorder parameter $l=\sigma_{\min}/(cE_F^2)$. All curves start at unity, recovering the Wiedemann--Franz law in the degenerate limit. For $l<1/4$ the ratio first dips \emph{below} unity before rising; for $l>1/4$ it rises monotonically. The ordering of the curves reverses in a narrow window near $t\approx0.6$.}
        \label{fig:lorenz}
    \end{figure}

    Figure~\ref{fig:lorenz} shows the corresponding Lorenz ratio. Every curve begins at $\mathcal{L}/\mathcal{L}_0=1$, as required: the Wiedemann--Franz law is exact in the degenerate limit $t\to0$, and its recovery here is a useful check on Eq.~(\ref{Lorenz_graphene}). At large $t$ the ratio grows without bound, reaching values above $3$ by $t=1.5$, so the law is strongly violated once the thermal window becomes comparable to $E_F$.

    The behaviour at small $t$ is more subtle, and is governed by the competition already identified below Eq.~(\ref{Lorenz_graphene}). Expanding for small $w=\pi^2t^2/3$ with $a=l+1$ gives

    \begin{equation} \label{lorenz_smallt}
        \frac{\mathcal{L}}{\mathcal{L}_0} \simeq
        1 + \frac{\tfrac{16}{5}a-4}{a^2}\,w + \mathcal{O}(w^2),
    \end{equation}

    so the initial slope changes sign at $a=5/4$, that is, at

    \begin{equation} \label{lorenz_threshold}
        l_c = \tfrac{1}{4}.
    \end{equation}
    
    For $l<l_c$ the subtraction term $-(4\pi^2t^2/3)/[(l+1)+\pi^2t^2/3]$, which represents the part of the energy current convected by the particle current, wins over the $7\pi^2t^2/5$ enhancement coming from the fourth moment $M_4$, and the Lorenz ratio is depressed below its Sommerfeld value before eventually rising. For $l>l_c$ the residual conductivity suppresses the thermopower strongly enough that the convective term never dominates, and the ratio increases monotonically from unity. It is visible in Fig.~\ref{fig:lorenz} as the curve $l=0.25$ leaving the origin with zero slope, separating the two families. The dip is shallow (at most $4\%$ in the clean limit) but it is a definite prediction of the model, and it has the opposite sign to the large-$t$ violation.

    The curves also reverse their ordering in a narrow window around $t\approx0.6$: below it a larger $l$ gives a larger Lorenz ratio, above it the ordering is inverted. The crossings are not exactly coincident, ranging from $t=0.577$ (for $l=0$ against $l=1$) to $t=0.664$ (for $l=0$ against $l=0.1$), but they are close enough to compress the whole family into a narrow waist. Table~\ref{tab:lorenz_data} summarises the relevant numbers.

    \begin{table}[h!]
        \centering
        \begin{tabular}{c|c|c|c|c}
            $l$ & behaviour at small $t$ & $t_{\min}$ & $(\mathcal{L}/\mathcal{L}_0)_{\min}$ & $\mathcal{L}/\mathcal{L}_0$ at $t=1.5$ \\
            \hline
            $0$    & dip below unity & $0.183776$ & $0.960000$ & $3.399742$ \\
            $0.1$  & dip below unity & $0.146089$ & $0.986909$ & $3.376392$ \\
            $0.25$ & zero slope      & $0$        & $1$        & $3.342171$ \\
            $0.5$  & monotonic rise  & $0$        & $1$        & $3.287191$ \\
            $1$    & monotonic rise  & $0$        & $1$        & $3.184373$
        \end{tabular}
        \caption{Small-$t$ behaviour, position and depth of the minimum, and endpoint value of the Lorenz ratio for the curves of Fig.~\ref{fig:lorenz}. The sign of the initial slope is set by Eq.~(\ref{lorenz_smallt}); the minimum lies at $t>0$ only for $l<1/4$.}
        \label{tab:lorenz_data}
    \end{table}

    \subsection{Electronic figure of merit and its universal maximum}\label{subsec:ZTe}

    We can also find the \textit{electronic-only} Figure of Merit $ZT_e$. In order to do this, we'll subsitute (\ref{Coeficients}) in the ordinary form for $ZT$ \cite{snyder_complex_2008,goldsmid_introduction_2016}.

    \begin{equation*}
        ZT_e = \frac{S^2\sigma}{\kappa_e}T = \frac{\left(-\frac{1}{eT}\,\frac{\mathcal{K}_1}{\mathcal{K}_0}\right)^2\left(\frac{2e^2}{h}\mathcal{K}_0\right)T}{\frac{2}{hT}\left(\mathcal{K}_2-\frac{\mathcal{K}_1^2}{\mathcal{K}_0}\right)}.
    \end{equation*}

    After some simplification we are left with

    \begin{equation}
        ZT_e = \frac{\mathcal{K}_1^2}{\mathcal{K}_2\mathcal{K}_0-\mathcal{K}_1^2}.
    \end{equation}

    This dimensionless combination of the three moments is the standard expression of the electronic figure of merit in linear-response theory; the same structure appears, for instance, in the exact Fermi--Dirac-integral analyses of Refs.~\cite{yadav_analytic_2019,nair_modeling_2026}. We emphasize that $ZT_e$ contains only the electronic thermal conductivity. The lattice contribution, which lowers the practical figure of merit, is reintroduced in Section 3. And then we substitute (\ref{Graphene_transport_numbers}) here, and after some simplification we obtain

    \begin{equation}
        ZT_e = \frac{(2\frac{\pi^2}{3}t^2)^2}{(l+1+\frac{\pi^2}{3}t^2)((l+1)\frac{\pi^2}{3}t^2+\frac{7\pi^4}{15}t^4)-(2\frac{\pi^2}{3}t^2)^2}.
    \end{equation}

    In order to simplify this, we make the transformation $w=\pi^2t^2/3$, then

    \begin{equation}
        ZT_e = \frac{(2w)^2}{(l+1+w)((l+1)w+21w^2/5)-4w^2},
    \end{equation}

    where we used that $\frac{7\pi^4}{15}t^4 = \frac{21}{5}w^2$. Now we expand the denominator, and by making algebra we obtain $(l+1+w)((l+1)w+21w^2/5)-4w^2 = w[(l+1)^2+(26(l+1)/5-4)w+21w^2/5]$, therefore, by returning the transformation we obtain.

    \begin{equation} \label{ZTe_graphene}
        ZT_e = \frac{4\frac{\pi^2}{3}t^2}{(l+1)^2+(\frac{26(l+1)}{5}-4)\frac{\pi^2}{3}t^2+\frac{7\pi^4}{15}t^4}.
    \end{equation}

    It is important to know at which values of $(l,t)$ the $ZT_e$ reaches the maximum value. In order to do that, we must find the critical points of $ZT_e$. We can notice that finding the maximum value of $ZT_e$ is equivalent to finding the minimum of $1/ZT_e$, so let's do that. We can express $1/ZT_e$ as

    \begin{equation}
        \frac{1}{ZT_e} = \frac{3}{4\pi^2} \left[ \frac{(l+1)^2}{t^2} + \frac{\pi^2}{3}\left(\frac{26(l+1)}{5} - 4\right) + \frac{7\pi^4t^2}{15} \right].
    \end{equation}

    $1/ZT_e$ is the sum of a term decreasing in $t^2$, a constant, and a term increasing in $t^2$, so a unique minimum exists, fixed by the balance of the first and last terms (this is the structure of the arithmetic--geometric mean inequality). Now, let's remember the definitions of $(l,t)$, which are in (\ref{Graphene_transport_abreviations}). We can see that $t$ is defined in terms of the temperature, so we must find the extrema when $t$ is the one that changes. Defining $g(t)=1/ZT_e$ and then taking the derivative with respect to $t$ leave us with

    \begin{equation}
        g^\prime(t) = \frac{3}{4\pi^2} \left[ -\frac{2(l+1)^2}{t^3} + \frac{14\pi^4t}{15} \right].
    \end{equation}

    We set this equal to zero and then

    \begin{equation}
        \frac{14\pi^4t^4}{15} - 2(l+1)^2 = 0 \:\:\: \Longrightarrow \:\:\: t_{\mathrm{opt}} = \frac{\sqrt{l+1}}{\pi}\left(\frac{15}{7}\right)^{1/4}. 
    \end{equation}

    Now, we use this result to calculate $ZT_e^{\max}$, we obtain

    \begin{equation} \label{ZTe_max}
        ZT_e^{\max}(l) = \frac{4\sqrt{15/7}/3}{2(l+1) + \sqrt{15/7}\left[26(l+1)/15 - 4/3\right]}.
    \end{equation}

    Two features of this result are worth highlighting. First, $ZT_e^{\max}$ decreases monotonically with $l$. The residual conductivity $\sigma_{\min}$ contributes to charge and heat transport but not to the particle--hole asymmetry, so it can only degrade the thermoelectric response. The best possible case is therefore the clean limit $l\to0$, for which Eq.~(\ref{ZTe_max}) gives the universal, parameter-free values

    \begin{equation}
        t_{\mathrm{opt}}(l=0) = \frac{1}{\pi}\left(\frac{15}{7}\right)^{1/4} \approx 0.3851
        \quad\Longleftrightarrow\quad
        E_F^{\mathrm{opt}} \approx 2.60\,k_BT,
        \qquad
        ZT_e^{\max}(l=0) \approx 0.7549.
    \end{equation}

    Second, this ceiling is intrinsic: within the quadratic conductivity model, no choice of doping or temperature can push the purely electronic figure of merit of graphene above $\approx0.7549$. We verified both numbers by direct numerical maximization of $ZT_e$ computed from the defining integrals (\ref{Graphene_transport_integrals}) at 30-digit precision, obtaining $ZT_e^{\max}=0.754890704787\ldots$ at $E_F/k_BT=2.59657\ldots$, in exact agreement with Eq.~(\ref{ZTe_max}). The optimal doping $E_F^{\mathrm{opt}}\approx2.6\,k_BT$ falls in the intermediate-degeneracy window singled out by general analyses of thermoelectric optimization \cite{mahan_sofo_1996,yadav_analytic_2019}, and lies precisely in the regime where the leading-order Mott formula is least reliable---which is why the exact terminated series is required to locate it correctly. \\

    We can also compute $ZT_e$ for different values of $l$, by doing that we obtain the following results.

    \begin{figure}[h!]
        \centering
        \includegraphics[width=0.75\textwidth]{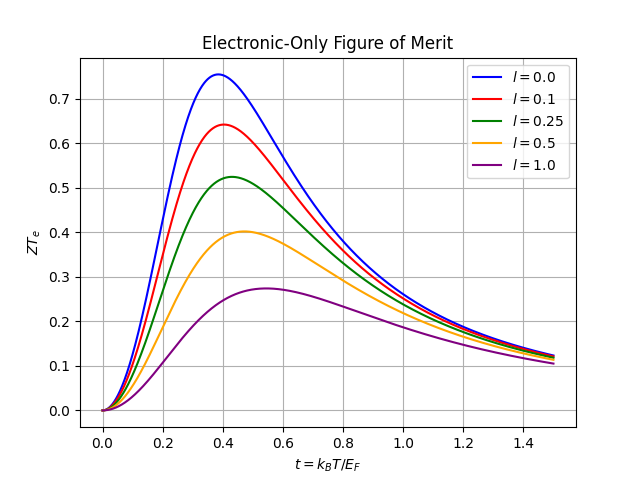}
        \caption{Electronic figure of merit $ZT_e$ of Eq.~(\ref{ZTe_graphene}) as a function of $t=k_BT/E_F$, for five values of the disorder parameter $l=\sigma_{\min}/(cE_F^2)$. Each curve rises from zero, passes through a single maximum, and decays; the maximum is highest and leftmost in the clean limit $l=0$, and all curves merge at large $t$.}
        \label{fig:ZTe}
    \end{figure}

    Figure~\ref{fig:ZTe} shows $ZT_e$ for $l=0,\,0.1,\,0.25,\,0.5,\,1$. The maximum of each curve sits at $t_{\mathrm{opt}}=\sqrt{l+1}\,(15/7)^{1/4}/\pi$ from Eq.~(\ref{ZTe_max}); in the clean limit this is $t_{\mathrm{opt}}=0.385122$, i.e. $E_F\simeq2.60\,k_BT$, where $ZT_e^{\max}=0.754891$. At large $t$ every curve approaches the common, $l$-independent asymptote $ZT_e\to20/(7\pi^2t^2)$. Three features are worth emphasising.

    First, every curve vanishes at both ends, and for the same reason in each case. As $t\to0$ the numerator of Eq.~(\ref{ZTe_graphene}) vanishes as $t^2$, the transport window becomes so narrow that the conductivity is effectively particle--hole symmetric about $E_F$, $\mathcal{K}_1\to0$, and the thermopower with it. As $t\to\infty$ the quartic term $7\pi^4t^4/15$ dominates the denominator and $ZT_e$ falls off as $20/(7\pi^2t^2)$, independently of $l$ --- visible in Fig.~\ref{fig:ZTe} as the merging of all five curves on the right-hand side. The maximum is therefore the result of a genuine competition between these two limits rather than of any fine-tuning.

    Second, the position of the maximum moves to higher $t$ as $l$ increases, following $t_{\mathrm{opt}}\propto\sqrt{l+1}$, while its height falls. Expressed in terms of the more familiar variable $E_F/k_BT=1/t$, the optimal doping decreases with disorder, from $2.60\,k_BT$ in the clean limit to $1.84\,k_BT$ at $l=1$. The residual conductivity $\sigma_{\min}$ adds to $\mathcal{K}_0$ and $\mathcal{K}_2$ but not to $\mathcal{K}_1$, since it is energy independent and therefore carries no particle--hole asymmetry. It dilutes the thermoelectric response without contributing to it, and the optimum shifts to compensate.

    Third, and most importantly, the clean-limit curve is an envelope. Since $ZT_e^{\max}(l)$ decreases monotonically in $l$ and $l\geq0$ by construction, the value $ZT_e^{\max}=0.754891$ is an upper bound for the model as a whole: within the quadratic conductivity model, no combination of doping and temperature can produce a purely electronic figure of merit above this number. Table~\ref{tab:ZTe_optima} collects the optima for the values of $l$ plotted.

    \begin{table}[h!]
        \centering
        \begin{tabular}{c|c|c|c}
            $l$ & $t_{\mathrm{opt}}$ & $E_F^{\mathrm{opt}}/k_BT$ & $ZT_e^{\max}$ \\
            \hline
            $0$    & $0.385122$ & $2.59658$ & $0.754891$ \\
            $0.1$  & $0.403919$ & $2.47574$ & $0.642193$ \\
            $0.25$ & $0.430580$ & $2.32245$ & $0.524695$ \\
            $0.5$  & $0.471676$ & $2.12010$ & $0.402084$ \\
            $1$    & $0.544645$ & $1.83606$ & $0.274018$
        \end{tabular}
        \caption{Position and height of the maximum of $ZT_e$, from Eq.~(\ref{ZTe_max}), for the curves of Fig.~\ref{fig:ZTe}. The clean-limit entry $ZT_e^{\max}=0.754891$ at $E_F^{\mathrm{opt}}=2.59658\,k_BT$ is the universal ceiling of the model.}
        \label{tab:ZTe_optima}
    \end{table}

    \subsection{Comparison with existing treatments of the same model}

    It is useful to state precisely what the closed forms of this section add to the two treatments already available for this problem.

    Antidormi and Cummings \cite{antidormi_optimizing_2021} evaluate the same integrals $\mathcal{K}_j$ for the same conductivity model by direct numerical quadrature, and use the results to optimise photothermoelectric device performance. Our Eqs.~(\ref{Seebeck_graphene}), (\ref{Lorenz_graphene}) and (\ref{ZTe_graphene}) are, within that model, algebraically identical to what such a quadrature returns---this is the content of the termination argument---so the two approaches never disagree. What changes is what can be said. Because the closed forms are elementary, they can be differentiated with respect to $t$ and inspected as functions of $l$, which yields three statements that quadrature alone does not supply: the saturation of $|S|$ at $|S|_{\max}=(k_B/e)\pi/\sqrt{3(l+1)}$, which converts their numerical observation that the Mott formula overestimates $S$ in high-mobility samples into an explicit mechanism and an explicit bound; the threshold $l_c=1/4$ separating samples whose Lorenz ratio dips below the Sommerfeld value from those whose ratio rises monotonically, which is a qualitative distinction invisible in a plot of a single sample; and the parameter-free ceiling $ZT_e^{\max}=0.754891$ together with the scaling $t_{\mathrm{opt}}\propto\sqrt{l+1}$, which bounds the whole model at once rather than one parameter set at a time. Their empirical remark that the Wiedemann--Franz law holds well for $E_F>2k_BT$ is likewise recovered and quantified by Eq.~(\ref{Lorenz_graphene}).

    The polylogarithm and Lambert-$W$ approach of Refs.~\cite{yadav_analytic_2019,nair_modeling_2026} is exact for a broader class of transport functions, since it evaluates the Fermi--Dirac integrals themselves rather than expanding a transport function about $\mu$. The present method is narrower in scope---it requires $G(E)$ to be polynomial for the series to terminate---but within that scope it returns elementary rational expressions in $t$ and $l$ rather than special functions of a reduced chemical potential that must be obtained by transcendental inversion. The two are therefore complementary: the Lambert-$W$ route handles arbitrary scattering exponents at the cost of special-function inversion, whereas the terminating zeta series handles the quadratic Dirac-material case with no special functions at all, and consequently delivers optima in closed form. The exact identity $S_{\mathrm{extreme}}\,t_{\mathrm{opt}}=-k_B/e$ of Eq.~(\ref{eq:seebeck_extremum}) is an example of a relation that is immediate in the present formulation.

    \section{Generalized Figure of Merit with Radiative Thermal Conductivity}

    \textit{The expressions in this section are an ansatz: they combine the generalized Seebeck coefficient $S_{\mathrm{GEN}}$ derived above with the standard definition of $ZT$ and the literature Rosseland radiative-conductivity formula. No new derivation of $ZT_{\mathrm{GEN,rad}}$ from a transport equation is claimed here.}

    \subsection{Definition and the generalized Seebeck substitution}

    The figure of merit condenses the competition between the useful thermoelectric response ($S^2\sigma$, the power factor) and the parasitic heat leakage ($\kappa$) into a single dimensionless number, and it governs the maximal conversion efficiency of any thermoelectric device \cite{goldsmid_introduction_2016}. Raising it has driven decades of materials research---through band-structure and nanostructure engineering \cite{hicks_effect_1993,dresselhaus_new_2007,majumdar_thermoelectricity_2004,heremans_when_2013,mao_size_2016,szczech_enhancement_2011}, the exploration of new material families \cite{hebert_searching_2016,shaabani_design_2018}, device-level optimization \cite{newbrook_mathematical_2020}, and, more recently, machine-learning-assisted discovery \cite{barua_recent_2025,bai_designing_2026,bos_roadmap_2026}. The usual thermoelectric figure of merit is defined as \cite{snyder_complex_2008,goldsmid_introduction_2016}

    \begin{equation}
        ZT=\frac{S^2\sigma T}{\kappa},
    \end{equation}

    where \(S\) is the Seebeck coefficient, \(\sigma\) is the electrical conductivity, \(T\) is the absolute temperature and \(\kappa\) is the thermal conductivity. Since in the present work the Seebeck coefficient is written in its generalized form, we replace

    \begin{equation}
        S \longrightarrow S_{\mathrm{GEN}}.
    \end{equation}

    From the previous derivation of the generalized Mott formula, we get a generalized electrical conductivity and Seebeck Coefficient, therefore, the first direct generalization of the thermoelectric figure of merit is

    \begin{equation}
        ZT_{\mathrm{GEN}} = \frac{S_{\mathrm{GEN}}^2\sigma}{\kappa}T.
    \end{equation}

    The substitution is conceptually the same one made in analytic studies of $ZT$ based on exact Fermi--Dirac integrals \cite{yadav_analytic_2019,nair_modeling_2026}. The leading-order (Mott/Wiedemann--Franz) transport coefficients are replaced by their all-orders counterparts, so that the figure of merit remains meaningful outside the strictly degenerate regime.

    \subsection{The radiative channel in the Rosseland diffusion limit}

    Now we include the thermal contribution associated with radiation. Interest in the interplay between photons and thermoelectric transport is not purely academic, photon fields can actively modify thermoelectric currents in cavity-coupled nanostructures \cite{abdullah_cavity_2016,abdullah_photon_2019}, illumination can enhance the power factor of real thermoelectric films \cite{lv_photoinduced_2015}, photonic (radiative) contributions can masquerade as an apparent Seebeck response in plasmonic metals \cite{zhao2025photonic}, and scanning-photothermoelectric experiments use light itself as the local heat source that probes $S$ \cite{evans_thermoelectric_2020}. Here, however, we restrict ourselves to the simplest and best-established role of radiation: as an additional passive heat-transport channel. In an optically thick medium, radiative heat transfer can be approximated by a diffusion process. In this limit, the radiative heat flux has a Fourier-like form and radiative transfer can be represented by an effective thermal conductivity \cite{christen_radiative_2010,siegel_thermal_1992}:

    \begin{equation}
        \vec{q}_{\mathrm{rad}}=-\kappa_{\mathrm{rad}}\nabla T.
    \end{equation}

    The corresponding radiative thermal conductivity is \cite{christen_radiative_2010,siegel_thermal_1992}

    \begin{equation}
        \kappa_{\mathrm{rad}}(T) = \frac{16\sigma_{\mathrm{SB}}T^3}{3\sigma_F^{(\mathrm{eff})}},
    \end{equation}

    where \(\sigma_{\mathrm{SB}}\) is the Stefan-Boltzmann constant and \(\sigma_F^{(\mathrm{eff})}\) is the Rosseland mean absorption coefficient. The validity of this diffusion (Rosseland) approximation requires the photon mean free path $1/\sigma_F^{(\mathrm{eff})}$ to be short compared with the size of the region across which the temperature varies; in optically thin structures the radiative loss is instead a surface (boundary) effect and cannot be folded into a bulk conductivity \cite{siegel_thermal_1992,christen_radiative_2010}. The characteristic $T^3$ growth of $\kappa_{\mathrm{rad}}$ signals that the radiative channel, negligible at cryogenic and often at room temperature, can become competitive in high-temperature thermoelectric operation, which is precisely the regime targeted by waste-heat-recovery applications \cite{snyder_complex_2008,bos_roadmap_2026}. The total thermal conductivity is then written as

    \begin{equation}
        \kappa_{\mathrm{tot}} = \kappa_e+\kappa_{\mathrm{ph}}+\kappa_{\mathrm{rad}},
    \end{equation}

    where \(\kappa_e\) is the electronic thermal conductivity, \(\kappa_{\mathrm{ph}}\) is the phonon or lattice thermal conductivity, and \(\kappa_{\mathrm{rad}}\) is the radiative thermal conductivity. Writing the three channels as additive assumes that they act in parallel and independently, which is the standard decomposition in thermoelectric analysis \cite{goldsmid_introduction_2016,yadav_analytic_2019}. Substituting the Rosseland radiative-diffusion approximation, we obtain

    \begin{equation}
        \kappa_{\mathrm{tot}} = \kappa_e+\kappa_{\mathrm{ph}} + \frac{16\sigma_{\mathrm{SB}}T^3}{3\sigma_F^{(\mathrm{eff})}}.
    \end{equation}

    Therefore, the generalized figure of merit including radiative heat transport is

    \begin{equation}
        ZT_{\mathrm{GEN,rad}} = \frac{S_{\mathrm{GEN}}^2\sigma T}{\kappa_e + \kappa_{\mathrm{ph}} + \dfrac{16\sigma_{\mathrm{SB}}T^3}{3\sigma_F^{(\mathrm{eff})}}}.
    \end{equation}

    \subsection{Limiting regimes and compact form}

    This expression keeps the same structure as the usual thermoelectric figure of merit, but now the denominator contains an additional thermal channel. This means that radiation contributes as a heat-loss mechanism. If $\kappa_{\mathrm{rad}} \ll \kappa_e+\kappa_{\mathrm{ph}}$, then the usual non-radiative approximation is recovered:

    \begin{equation}
        ZT_{\mathrm{GEN,rad}} \approx \frac{S_{\mathrm{GEN}}^2\sigma T}{\kappa_e+\kappa_{\mathrm{ph}}}.
    \end{equation}

    On the other hand, if $\kappa_{\mathrm{rad}} \sim \kappa_e+\kappa_{\mathrm{ph}}$, then radiative heat transport must be included explicitly. It is also useful to define $\kappa_0 = \kappa_e+\kappa_{\mathrm{ph}}$ so that

    \begin{equation}
        ZT_{\mathrm{GEN},0} = \frac{S_{\mathrm{GEN}}^2\sigma T}{\kappa_0}.
    \end{equation}

    Then the radiative-corrected figure of merit can be written as

    \begin{equation}
        ZT_{\mathrm{GEN,rad}} = \frac{ZT_{\mathrm{GEN},0}}{1+\kappa_{\mathrm{rad}}/\kappa_0}.
    \end{equation}

    Using the Rosseland expression for \(\kappa_{\mathrm{rad}}\), this becomes

    \begin{equation}
        ZT_{\mathrm{GEN,rad}} = \frac{ZT_{\mathrm{GEN},0}}{1+\dfrac{16\sigma_{\mathrm{SB}}T^3}{3\sigma_F^{(\mathrm{eff})}\left(\kappa_e+\kappa_{\mathrm{ph}}\right)}}.
    \end{equation}

    This form shows explicitly how radiative heat transport reduces the figure of merit. The compact multiplicative structure is convenient in practice: the entire radiative correction enters through the single dimensionless ratio $\kappa_{\mathrm{rad}}/\kappa_0$, so that any figure of merit computed without radiation (including the exact electronic $ZT_e$ of Section 2, once the lattice channel is added) can be corrected a posteriori. Since $\kappa_{\mathrm{rad}}\propto T^3$ while $\kappa_0$ typically varies much more slowly with temperature, the correction factor decreases monotonically with $T$, shifting the temperature of maximal $ZT$ downward relative to the non-radiative prediction; a quantitative evaluation of this shift for specific material parameters, together with the validity limits of the Rosseland approximation in thin-film and two-dimensional geometries, is left for future work.

\section{Conclusions}

    We have resummed the Sommerfeld expansion of the linear-response transport coefficients into a single indexed series of Riemann zeta functions, and applied it to the standard conductivity model of graphene photothermoelectric devices. The principal results are the following.

    \begin{enumerate}
        \item \textbf{Every moment of the thermal broadening kernel is elementary.} The moments $M_{2i}=\int x^{2i}e^{x}(e^{x}+1)^{-2}dx$ evaluate in closed form through the Dirichlet eta function, $M_{2i}=2\,(2i)!\,(1-2^{1-2i})\zeta(2i)$, Eq.~(\ref{M_generalsolution}), a form that also reproduces $M_0=1$ when read with $\zeta(0)=-1/2$. The interchange of summation and integration required at two points in the derivation is justified by an explicit majorant through the Fubini--Tonelli theorem rather than by uniform convergence, which fails on the unbounded domain.

        \item \textbf{An all-orders Mott formula.} Writing $\sigma$ and $S\sigma$ as zeta series in the even and odd energy derivatives of the transport function gives the zeta-function Mott series, Eq.~(\ref{Generalized_Seebeck}), whose leading term is the textbook Mott formula and whose higher terms supply the corrections of relative order $(k_BT/\mu)^2$ and beyond.

        \item \textbf{A termination criterion that makes the result exact.} When the transport function is a polynomial the two series contain finitely many terms. The resulting expressions are then not improved approximations but algebraic identities, exact and non-perturbative in $k_BT/\mu$ at every temperature for which the underlying model applies. The standard graphene photothermoelectric conductivity $\sigma(E)=\sigma_{\min}+cE^{2}$ is quadratic and falls exactly into this class.

        \item \textbf{Closed forms in two dimensionless parameters.} The Seebeck coefficient, the Lorenz ratio and the electronic figure of merit reduce to elementary rational functions of $t=k_BT/E_F$ and $l=\sigma_{\min}/(cE_F^{2})$ alone, Eqs.~(\ref{Seebeck_graphene}), (\ref{Lorenz_graphene}) and (\ref{ZTe_graphene}). All three were verified against direct quadrature of the defining integrals at 30-digit precision.

        \item \textbf{Three consequences that follow only because the forms are elementary.} The thermopower saturates at $|S|_{\max}=(k_B/e)\,\pi/\sqrt{3(l+1)}$, which explains in closed form the overestimation of $S$ by the bare Mott formula reported numerically for high-mobility graphene, and satisfies the exact identity $S_{\mathrm{extreme}}\,t_{\mathrm{opt}}=-k_B/e$. The Lorenz ratio has a disorder threshold $l_c=1/4$ separating samples whose ratio dips below the Sommerfeld value from those whose ratio rises monotonically---two qualitatively different modes of Wiedemann--Franz violation of opposite sign. The electronic figure of merit has a parameter-free ceiling $ZT_e^{\max}\simeq0.7549$ at $E_F\simeq2.60\,k_BT$, attained in the clean limit, with the optimum shifting as $t_{\mathrm{opt}}\propto\sqrt{l+1}$ as disorder increases.
    \end{enumerate}

    The optimal operating point of the model lies at $E_F$ of order a few $k_BT$, which is precisely the intermediate-degeneracy window where the leading-order Mott formula is least trustworthy and where room-temperature graphene photodetectors actually operate. An exact treatment is therefore not a refinement of the theory at its margins but a requirement for locating the optimum at all. More generally, the termination criterion identifies a class of problems---any material whose transport function is polynomial in energy over the relevant window---for which the entire Sommerfeld hierarchy collapses to a finite calculation; Dirac and Weyl materials, whose densities of states are power laws, are the natural candidates.

    The framework inherits the scope of the Jonson--Mahan picture: independent carriers, relaxation-time or static-disorder scattering, and a transport function analytic about $\mu$. Strong inelastic scattering, phonon drag, and the interaction-driven hydrodynamic regime near charge neutrality lie outside it, as does the quadratic model itself very close to the Dirac point. The radiative channel of Section 3 is included at the level of an ansatz, by combining the generalized Seebeck coefficient with the literature Rosseland conductivity; a derivation of the coupled radiative--electronic problem from a transport equation, and an assessment of the validity of the diffusion approximation in two-dimensional and thin-film geometries, remain open. The most immediate extension is to transport functions of higher polynomial degree, where the series still terminates and the same closed-form program can be carried through.

   \section*{Acknowledgment}
	PCD thanks the Anusandhan National Research Foundation (ANRF), Government of India,
	for the award ANRF/PMP/2025 that supported his work. Sree Ram (S. R.) Valluri and Luis
	Daniel Villa Cortés gratefully acknowledge the financial support provided by MITACS for this
	summer research project and thank the Chair, Professor Graham Denham, the staff of Math Department
	and Proffessor Carol Jones, Faculty Relations, for facilitating the Mitacs intership of Luis Daniel
	Villa Cortés and thank all of them for providing the space and facilities necessary to carry out
	this research. We also thank Professor Tatyana Barron for her very valuable contributions and
	feedback on the development of this paper.

    \printbibliography   

\end{document}